# Observation of a pressure-induced transition from interlayer ferromagnetism to intralayer antiferromagnetism in $Sr_4Ru_3O_{10}$

H. Zheng[1], W.H. Song[1,2], J. Terzic[3], H. D. Zhao[1], Y. Zhang[1], Y. F. Ni[1], L. E. DeLong[4], P. Schlottmann[5] and G. Cao[1*]

[1]Department of Physics, University of Colorado at Boulder, Boulder, CO 80309

[2]Institute of Solid State Physics, Chinese Academy of Sciences, Hefei 230031, China

[3]National High Magnetic Field Laboratory, Tallahassee, FL 32306

[4]Department of Physics and Astronomy, University of Kentucky, Lexington, KY 40506

[5]Department of Physics, Florida State University, Tallahassee, FL 32306

## Abstract

$Sr_4Ru_3O_{10}$ is a Ruddlesden-Popper compound with triple Ru-O perovskite layers separated by Sr-O alkali layers. This compound presents a rare coexistence of interlayer (*c*-axis) ferromagnetism and intralayer (basal-plane) metamagnetism at ambient pressure. Here we report the observation of pressure-induced, intralayer ***itinerant antiferromagnetism*** arising from the interlayer ferromagnetism. The application of modest hydrostatic pressure generates an anisotropy that causes a flattening and a tilting of $RuO_6$ octahedra. All magnetic and transport results from this study indicate these lattice distortions diminish the *c*-axis ferromagnetism and basal-plane metamagnetism, and induce a basal-plane antiferromagnetic state. The unusually large magnetoelastic coupling and pressure tunability of $Sr_4Ru_3O_{10}$ makes it a unique model system for studies of itinerant magnetism.

[*] Email: gang.cao@colorado.edu

**I. Introduction**

$Sr_4Ru_3O_{10}$ is a "triple-layer" member of the Ruddlesden-Popper ruthenates, $(Ca, Sr)_{n+1}Ru_nO_{3n+1}$ (n = number of Ru-O perovskite layers/unit cell). Ruthenates feature extended *4d*-electron orbitals and comparable energy scales (and therefore competition) among Coulomb interactions, crystalline electric fields (CEF), spin-orbit interactions, p-d hybridization and spin-lattice coupling. The deformations and relative orientations of corner-shared $RuO_6$ octahedra determine the CEF level splitting and the electronic band structure, and hence the ground state. Consequently, the physical properties of ruthenates are highly sensitive to dimensionality and susceptible to perturbations such as the application of magnetic field, and/or pressure. These characteristics are well demonstrated by the contrasting physical properties of $Ca_{n+1}Ru_nO_{3n+1}$ and $Sr_{n+1}Ru_nO_{3n+1}$ (n = 1, 2, 3, ∞): The Ca compounds are on the verge of a metal-insulator transition and prone to antiferromagnetism (AFM) whose character changes with n. In contrast, the Sr compounds are metallic, and evolve from paramagnetism (n = 1, 2) to ferromagnetism (FM) (n = ∞) with increasing n [1-20]. It is therefore not surprising that these materials exhibit nearly every cooperative phase known in solids.

The triple-layered $Sr_4Ru_3O_{10}$ (n = 3) [12] is precariously positioned on the borderline separating the ferromagnet $SrRuO_3$ (n = ∞) [5] and the field-induced metamagnet $Sr_3Ru_2O_7$ (n = 2) [9,10], and displays complex phenomena ranging from tunneling magnetoresistance and quantum oscillations [21] to a switching effect [15]. However, the most distinct, intriguing hallmark of $Sr_4Ru_3O_{10}$ is its seemingly contradictory magnetic behavior: for a magnetic field *along the c-axis* (perpendicular to the layers), it exhibits itinerant FM with a Curie temperature $T_C$ = 105 K and a saturation moment greater than 1.0 $\mu_B$/Ru (**Fig.1a**); on the other hand, when the magnetic field is applied *within the basal plane* it features a pronounced peak in the magnetization near $T_M$ = 50 K (**Fig.1a**) and a sharp metamagnetic transition near $H_c$ = 2.5 T (**Fig.1b**) [12]. This situation recalls the metamagnetic transitions out of Stoner exchange-enhanced paramagnetism in Sr-based Ruddlesen-Popper compounds [22-26]. The coexistence of the interlayer FM and the intralayer metamagnetism is not anticipated from simple theoretical arguments [22, 23, 27]. A two-dimensional, tight-binding electron gas has a logarithmic divergence in the density of states which, depending on the position of the Fermi level, can yield FM, metamagnetism and a quantum critical point by varying applied pressure [26]. However, $Sr_4Ru_3O_{10}$ is not strictly two-dimensional and a suitable model must be adapted accordingly; whereupon $Sr_4Ru_3O_{10}$ then provides an opportunity to study the interplay of itinerant FM, AFM and metamagnetism.

The peculiar behavior of $Sr_4Ru_3O_{10}$ (n = 3) has drawn considerable attention in recent years. Discussions have mainly focused on the relationship between the interlayer FM and the intralayer metamagnetism below $T_M$, the onset of the metamagnetic state. A Raman study explores the spin-lattice coupling as a function-of temperature, magnetic field and pressure, and informs a microscopic description of structural and magnetic phases [14]. Specifically, magnetic-field-induced changes in the phonon spectra reveal spin-reorientation transitions and strong magnetoelastic coupling below $T_M$ [14]. Neutron scattering studies [28, 29] indicate a collinear FM state aligned along the *c*-axis with no detectable spin canting toward the basal plane. The metamagnetic transition was attributed to magnetic domains [28] that disappear above $T_M$ according to magneto-optical imaging and scanning Hall probe measurements [30]. On the other hand, the magnetic-field dependence of the bulk heat capacity does exhibit abrupt changes at the metamagnetic transition $H_c$, suggesting the metamagnetic behavior is not due to domains [31]. In addition, a rapid increase in the *c*-axis lattice parameter below $T_M$ at ambient conditions is observed [29], signaling a critical role for strong spin-lattice coupling in determining the magnetic state. A similar conclusion is also drawn from studies of Raman scattering [14] and Ca- and La-doping behavior of $Sr_4Ru_3O_{10}$ [32]. A recent magnetic and transport study of nanoscale flakes (30 – 350 nm) of $Sr_4Ru_3O_{10}$ revealed a drastic size effect in the evolution from *c*-axis FM to a basal-plane AFM state [33]. In short, it is increasingly clear that spin-lattice coupling is critical to the apparent coexistence of the interlayer FM and the intralayer metamagnetism.

Application of pressure is a powerful tool for tuning lattice properties without introducing disorder, and should provide insights into the complex behavior of $Sr_4Ru_3O_{10}$. This work reports measurements of the magnetic and transport properties of bulk single-crystal $Sr_4Ru_3O_{10}$ at pressures up to 25 kbar in applied magnetic fields up to 14 T. Applied pressure can tune interactions over an energy range of ~ 10 meV (e.g., 10 meV/Å$^3$ ≈ 16 kbar). Here, we report a rapid evolution from *c*-axis itinerant FM at ambient pressure to basal-plane itinerant AFM at pressures near 25 kbar, accompanied by a vanishing magnetic anisotropy that promotes the metamagnetic state at ambient pressure. We find the application of hydrostatic pressure to $Sr_4Ru_3O_{10}$ generates an anisotropic effect that induces a remarkable flattening and tilting of the $RuO_6$ octahedra, which depresses the *c*-axis FM and metamagnetism, along with the exchange anisotropy, and induces a basal-plane AFM state with pressure. This study indicates an unusually large magnetoelastic coupling operates in $Sr_4Ru_3O_{10}$, which makes it an ideal model system for studies of competing itinerant FM, AFM and metamagnetism.

**II. Experimental details**

Single crystals of $Sr_4Ru_3O_{10}$ were grown using a self-flux method from off-stoichiometric quantities of $RuO_2$, $SrCO_3$, and $SrCl_2$. The average sample size is approximately 1 x 1 x 0.4 $mm^3$. Measurements of crystal structures were performed using a Bruker D8 Quest ECO single-crystal diffractometer equipped with a PHOTON 50 CMOS detector. Chemical analyses of the samples was performed using a combination of a Hitachi MT3030 Plus Scanning Electron Microscope and an Oxford Energy Dispersive X-Ray Spectroscopy (EDX). Standard four-lead measurements of the electrical resistivity were carried out using a Quantum Design (QD) Dynacool PPMS System equipped with a 14-Tesla magnet. Magnetic properties were measured using a QD MPMS-7 SQUID Magnetometer. Two hydrostatic pressure cells compatible with the QD instruments were used for measurements of electrical resistivity (up to 27 kbar) and magnetization (up to 13 kbar).

**III. Results and discussion**

The orthorhombic structure of $Sr_4Ru_3O_{10}$ is a slightly distorted cubic structure consistent with a *Pbam* space group, and room-temperature lattice parameters a = 5.4982 Å, b = 5.4995 Å and c = 28.5956 Å. One important structural detail is that in the outer two perovskite layers, the $RuO_6$ octahedra are rotated 5.25° about the *c*-axis, whereas in the middle layer they are rotated 10.6° about the *c*-axis in the ***opposite direction*** (see **Fig.1 Inset**) [12]. This structural feature has significant implications for the spin configuration and physical properties discussed below, due to the action of strong spin-lattice coupling.

It is established that the spins are ferromagnetically aligned along the *c*-axis in Sr$_4$Ru$_3$O$_{10}$ at ambient conditions, effectively forming FM chains along the *c*-axis [14, 28, 29; note these studies differ on whether or not spin canting exists at ambient pressure]. Our magnetic data indicate that application of a modest pressure readily destabilizes the *c*-axis FM state, and fosters emergent AFM correlations with spins primarily aligned within the basal plane below T$_M$, as illustrated in **Fig. 2** (note the applied magnetic field is merely 0.1 T). The pressure-induced change in T$_M$ is clearly identified by the shift of a corresponding peak in the basal-plane magnetization M$_{ab}$, and indicates an astonishing fourfold enhancement of M$_{ab}$ as the applied pressure P increases from 0 kbar to 10 kbar (**Fig. 2a**), whereas the *c*-axis magnetization M$_c$ undergoes a comparable reduction (**Fig. 2b**). A peak in M$_c$ emerges at around P = 8 kbar, and becomes well-defined at P = 10 kbar. The occurrence of peaks in both M$_{ab}$ and M$_c$ at P = 10 kbar serve as clear signatures of an AFM ordered state (**Fig. 2c**). Note that the observed anisotropy in M$_{ab}$ and M$_c$, which at P = 0 kbar is one order of magnitude (**Fig. 2c Inset**), almost vanishes near P = 10 kbar. In addition, the signature step in M$_c$ at T$_C$ decreases with P (see discussion below). Overall, applied pressure rapidly drives the magnetic state from a *c*-axis FM state toward a basal-plane AFM state below T$_M$. It is striking that a mere 10 kbar can cause such drastic changes in the magnetization and easy direction, which demonstrates an unusually strong magnetoelastic effect is at play in this material.

A Raman study of Sr$_4$Ru$_3$O$_{10}$ revealed a strong spin-phonon coupling of 5.2 cm$^{-1}$ below T$_M$ [14]. We infer that the pressure-induced AFM state is likely a result of a flattening of the RuO$_6$ octahedra, in which $d_{xy}$ orbitals have the lowest energy and are fully occupied, whereas the $d_{xz}$ and $d_{yz}$ orbitals are half-filled. Superexchange interactions mediated by electrons in the $d_{xz}$ and $d_{yz}$ orbitals favor an AFM state (**Fig. 2d**), according to electronic band structure calculations [6, 18, 33]. This scenario also explains our observation of an unusual increase in the *c*-axis resistivity $\rho_c$ with P at T > $T_C$, as shown in **Fig. 3**. The basal-plane resistivity $\rho_{ab}$ changes only slightly with P, showing a small decrease with P at higher temperatures, presumably as a result of band broadening (**Fig. 3a**). The slight change in $\rho_{ab}$ implies that soft-phonon and spin-disorder scattering within the basal plane are largely unaffected under the conditions studied here. In sharp contrast, the *c*-axis resistivity $\rho_c$ increases by a factor of two at T > $T_C$ (**Fig. 3b**); in particular, the ratio of $\rho_c/\rho_{ab}$ at 300 K rises from 3.8 at ambient pressure to 7.2 at 25 kbar (**Fig. 3a inset**). This behavior suggests applied pressure induces a tilting of the $RuO_6$ octahedra along the *c*-axis, which reduces the Ru-O-Ru bond angle from 180º (**Fig. 3d**), which, in turn, reduces the overlap of *p*- and *d*-orbitals. This would explain the significant increase in $\rho_c$ for T > $T_C$ (**Fig. 3b**), and is also consistent with an anomalous pressure dependence of the 380 cm$^{-1}$ $B_{1g}$ phonon mode at low temperatures, which is attributed to a buckling of the $RuO_6$ octahedra [14]. Interestingly, the magnitude of $\rho_c$ for T < $T_M$ is much less affected by applied pressure (**Fig. 3b**). The rapidly reduced $\rho_c$ below $T_M$ signifies a strengthened overlap of $d_{xz}/d_{yz}$ orbitals, and, more importantly, the existence of long-range magnetic order at P = 25 kbar; this is because electrical transport is intimately coupled to the magnetism, and long-range order significantly reduces both phonon and spin scattering.

Also of interest is the temperature dependence of $\rho_c \sim T^{\alpha}$ at low temperatures (1.8-10 K), where the exponent $\alpha$ changes significantly from near 2 at ambient pressure to 3/2 (see **Fig. 3b Inset**), which suggests a dominance of AFM spin fluctuations and a breakdown of the Fermi liquid model [22]. As shown in **Fig. 3c** the value of $\alpha$ changes more rapidly near P = 10 kbar, which may mark an onset of a more isotropic itinerant AFM state [22]. The change in exponent α is correlated with changes of $T_M$, which increases at P ≥ 10 kbar, while $T_C$ steadily decreases over the same range (**Fig. 3b** shows $\rho_c$ at a few representative pressures). The opposite response of $T_M$ and $T_C$ to P further confirms that the basal-plane AFM state becomes more energetically favorable than the *c*-axis FM state with increasing distortion of the $RuO_6$ octahedra with increasing P. The data in **Fig. 3** imply that the emergent basal-plane AFM state is dominant at P ≥ 10 kbar.

The evolution of both the basal-plane and *c*-axis isothermal magnetizations at low temperatures also indicate that applied pressure enhances $M_{ab}$ and weakens $M_c$, as shown in **Fig. 4**. The critical field of the metamagnetic transition, $H_c$, decreases with increasing P, indicative of an increasingly weakened magnetic anisotropy (**Fig. 4a**). Note that magnetic field applied along the *c*-axis helps elongate the $RuO_6$ octahedra along the *c*-axis [14], thus enhancing $M_c$. This effect competes with applied pressure that tends to compress the $RuO_6$ octahedra. The result of the two competing effects may explain why $M_c$ changes only modestly with P (**Fig. 4b**). Moreover, both $M_{ab}$ and $M_c$ exhibit a sizable hysteresis effect at ambient pressure (**Fig.1**) [12], but this hysteresis almost vanishes at P = 10 kbar (not shown), consistent with a weakened FM state.

The system exhibits a large negative magnetoresistivity ρ(H) with an overall reduction of up to 80% (**Fig. 5**; note that H is applied within the basal plane for both $\rho_{ab}$ and $\rho_c$). We propose that the basal-plane resistivity $\rho_{ab}(H)$ rises initially due to the canted AFM state, and then drops abruptly when H is strong enough to align the spins in a collinear fashion, which reduces spin scattering. Specifically, $\rho_{ab}$ shows two peaks near two critical fields, $H_c$ and $H_{c2}$, respectively. The peak at $H_c$ marks the metamagnetic transition that signals a spin-flip (presumably in the middle layer) that aligns the middle layer spins with those in the outer layers; therefore all spins, although canted, are approximately polarized along with the direction of H, which reduces spin scattering, and explains the abrupt drop in $\rho_{ab}$ near $H_c$ (**Fig. 5a**). This is then followed at higher fields by another drop in $\rho_{ab}$ at $H_{c2}$ (**Figs. 5a** and **5c**), which is already apparent but not well-defined in $M_{ab}$ in **Fig. 4a**. It indicates an additional spin alignment which eventually diminishes the spin canting and further reduces spin scattering. The evolution of the spin configuration with H is schematically illustrated in **Fig. 5d**. Note that $\rho_c$ drops even more sharply near $H_c$, but exhibits no anomaly at $H_{c2}$; it instead increases linearly with H when $H > H_c$ (**Fig. 5b** and **5c**). Since the direction of H is perpendicular to the direction of electrical current, the linear rise of $\rho_c$ with H could be a result of the familiar deflection (orbital magnetoresistance) of electrons by the Lorentz force. On the other hand, this linear field-dependence is strikingly similar to that observed in $Ca_3Ru_2O_7$, which is attributed to an orbital order that strengthens with increasing magnetic field, which hinders electron hopping [36, 37]. The two critical fields $H_c$ and $H_{c2}$ rapidly decrease with increasing P and eventually vanish at P = 25 kbar, where $\rho_{ab} \sim H^2$ ($\rho_c$ behaves similarly above 1 T). This trend is also evident in $M_{ab}$ and $M_c$ in **Fig. 4**. If an orbital order does indeed exist here, it is substantially weakened at 25 kbar (**Fig. 5b**).

It needs to be pointed out that the field dependences of both $\rho_{ab}$ and $\rho_c$ bear a strong resemblance to the *bulk spin-valve effect* observed in bilayered $Ca_3(Ru_{1-x}Cr_x)_2O_7$, which originates from inhomogeneous exchange coupling and soft and hard bilayers having antiparallel spin alignments [38]. Consistently, the distinct behavior demonstrated by $\rho_{ab}$ and $\rho_c$ (including somewhat different $H_c$) (**Figs. 5a-5c**) implies a strong exchange anisotropy that must arise from the competition between FM and AFM correlations. The anisotropy and spin-valve effect are clearly highly susceptible to the applied pressure and eventually vanish when P is greater than 25 kbar.

### IV. Conclusions

A temperature-pressure phase diagram can be generated using the results of this study, as shown in **Fig. 6**. Quasi-hydrostatic pressure surprisingly generates an anisotropic flattening of the $RuO_6$ octahedra, and reduces the Ru-O-Ru bond angle along the *c*-axis. This is implied by the considerable increase in $\rho_c$ above $T_C$ (**Fig. 3b**). The lattice distortions destabilize the *c*-axis FM state, and foster a basal-plane AFM state (**Figs. 2** and **4**). As a result, $T_C$ decreases at a rate $dT/dP \approx -1$ K/kbar; whereas $T_M$, which defines the onset of the AFM state, decreases initially and then rises for $P \geq 10$ kbar (**Figs. 2** and **3**). Indeed, the rapid change in the temperature dependence of $\rho_c$ (and $\rho_{ab}$) from $T^2$ to $T^{3/2}$ below 10 K near P = 10 kbar marks a crossover from the *c*-axis FM state to a predominantly basal-plane AFM state (**Fig. 3c**). Nevertheless, the opposite pressure responses of $T_C$ and $T_M$ point to a merging of the two magnetic states at $P \geq 25$ kbar, at which a collinear, itinerant AFM state is presumably fully established (**Figs. 5** and **6**). The existence of this pressure-induced long-range order at 25 kbar is strongly indicated by the abrupt drop in $\rho_c$ below $T_M$ (**Figs. 3a inset** and **3b**). We note that the attainment of a collinear antiferromagnetic state out of a low-pressure canted state should also generate a strongly varying anomalous (topological) Hall effect, due to variations in the scalar spin chirality [39].

It is remarkable that the FM state with $T_C$ = 165 K in the sister compound $SrRuO_3$ decreases with pressure at a slower rate $dT/dP \approx -0.68$ K/kbar, and only vanishes in a much higher pressure range of 170 to 340 kbar, where a paramagnetic state emerges [35]. This sharply contrasts with the high tunability offered by pressure applied in $Sr_4Ru_3O_{10}$, and highlights the rich physics available for study in this peculiar layered magnet.

**Acknowledgements**: GC is thankful to Dr. Mingliang Tian for useful discussions and to the High Magnetic Field Laboratory, Chinese Academy of Sciences for the hospitality during which part of this paper was drafted. This work was supported by NSF grant DMR-1712101; research of LED is supported by NSF Grant No. DMR-1506979.

**Figure Captions:**

**Fig.1**. **(a)** Temperature dependence of the magnetization M at $\mu_oH = 0.01$ T, and **(b)** Isothermal magnetization M(H) at T =1.8 K for both the basal-plane $M_{ab}$ and the *c*-axis $M_c$ at ambient pressure. **Inset**: Schematic of the triple-layered crystal structure; the curved arrows indicate the rotation of the $RuO_6$ octahedra. Note that the data in this figure were taken earlier at ambient pressure without the pressure cell, and they are presented here only to serve as part of the introduction of the title material.

**Fig.2**. Temperature dependence of the magnetization M at $\mu_oH = 0.1$ T for **(a)** the basal plane $M_{ab}$, **(b)** the *c*-axis $M_c$ at representative pressures and **(c)** $M_{ab}$ and $M_c$ at P = 10 kbar for comparison. **Inset**: $M_{ab}$ and $M_c$ at P = 0 kbar. **(d)** Schematic for pressure-induced changes in the spin configuration. Note that the magnitude of $M_{ab}$ (and $M_c$) rapidly increases (decreases) with P, which is highlighted by the two broad vertical arrows.

**Fig.3**. Temperature dependence of the electrical resistivity for **(a)** the basal plane $\rho_{ab}$, **(b)** the *c*-axis $\rho_c$ at representative pressures. Note the modest decrease in $\rho_{ab}$ and the considerable increase in $\rho_c$ with P, which is marked by the broad arrow. **(c)** The exponent $\alpha$ of $\rho_c \sim T^\alpha$ as a function of pressure P. Note that the shaded area marks the rapid change in $\alpha$. **(d)** Schematic for pressure-induced changes in the $RuO_6$ octahedra.

**Fig. 4**. The isothermal magnetization at T = 2 K for **(a)** the basal-plane $M_{ab}$ for H ∥ basal plane and **(b)** the *c*-axis $M_c$ for H ∥ *c*-axis at a few representative pressures.

**Fig.5**. Magnetic-field dependence of the electrical resistivity for **(a)** the basal plane $\rho_{ab}$, **(b)** the *c*-axis $\rho_c$ at representative pressures, and **(c)** $\rho_{ab}$ and $\rho_c$ and $M_{ab}$ (right scale) at P = 8 kbar for comparison. **(d)** Schematic for field-induced changes in the spin configuration.

Note that the two broad horizontal arrows in Fig.4b are to highlight the rapid decrease in $H_c$ with P, and the vertical dashed line in Fig.4c indicates $H_c$.

**Fig.6**. A T-P phase diagram generated based on the magnetic and transport results. Note that $T_C$ and $T_M$ appear to merge as the antiferromagnetic state is fully developed at P ≥ 25 kbar. Note that the shaded areas near 10 kabr marks a crossover from the *c*-axis FM state to a predominately basal-plane AFM state.

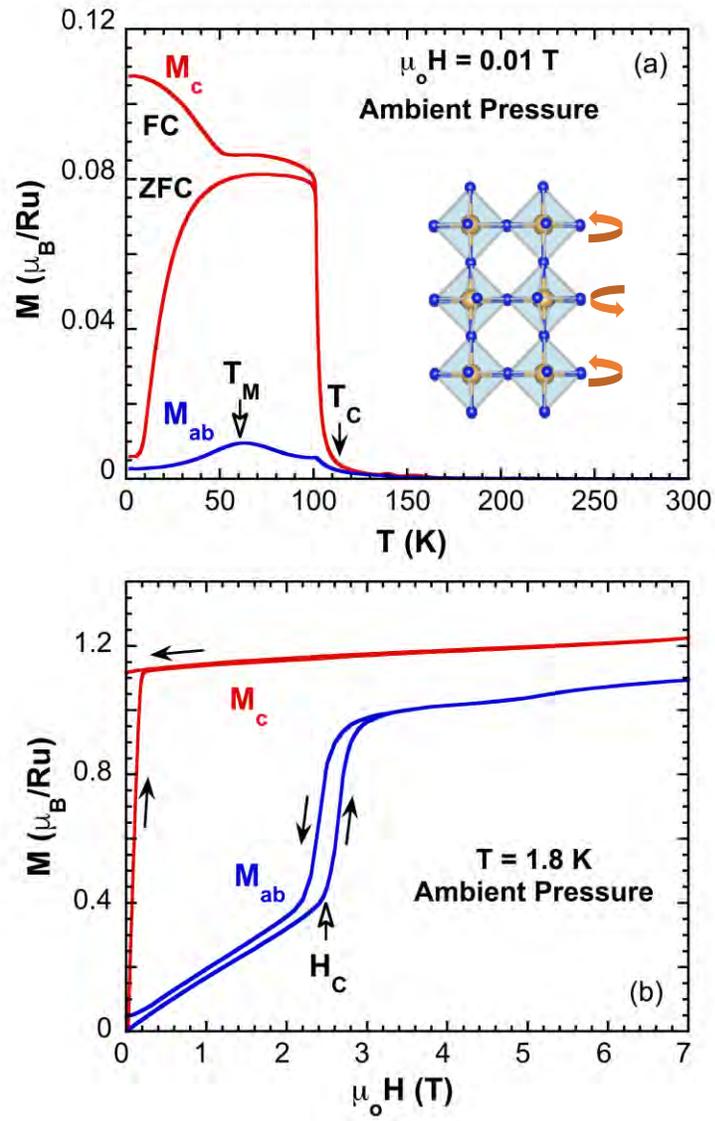

**Fig. 1**

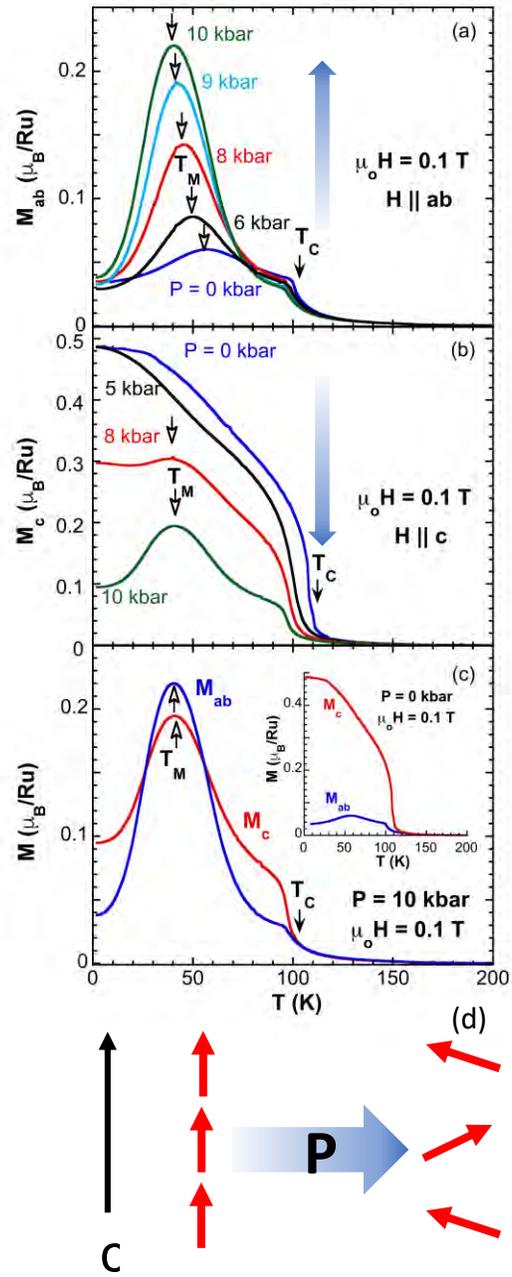

**Fig. 2**
19

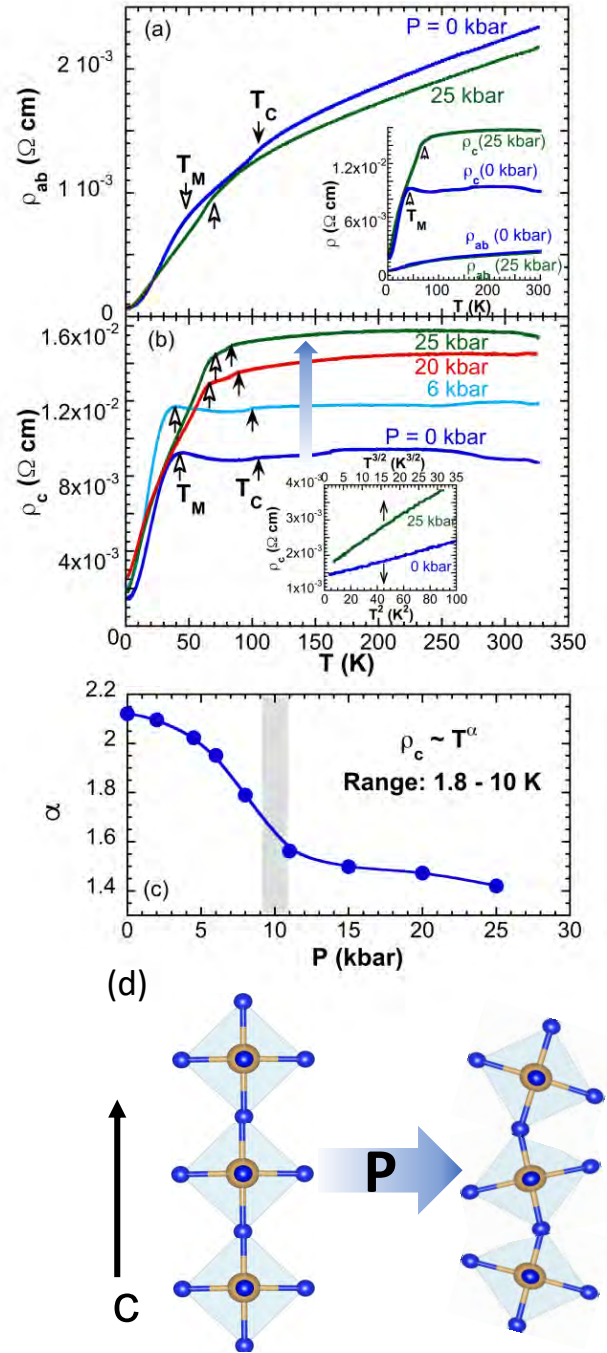

**Fig. 3**

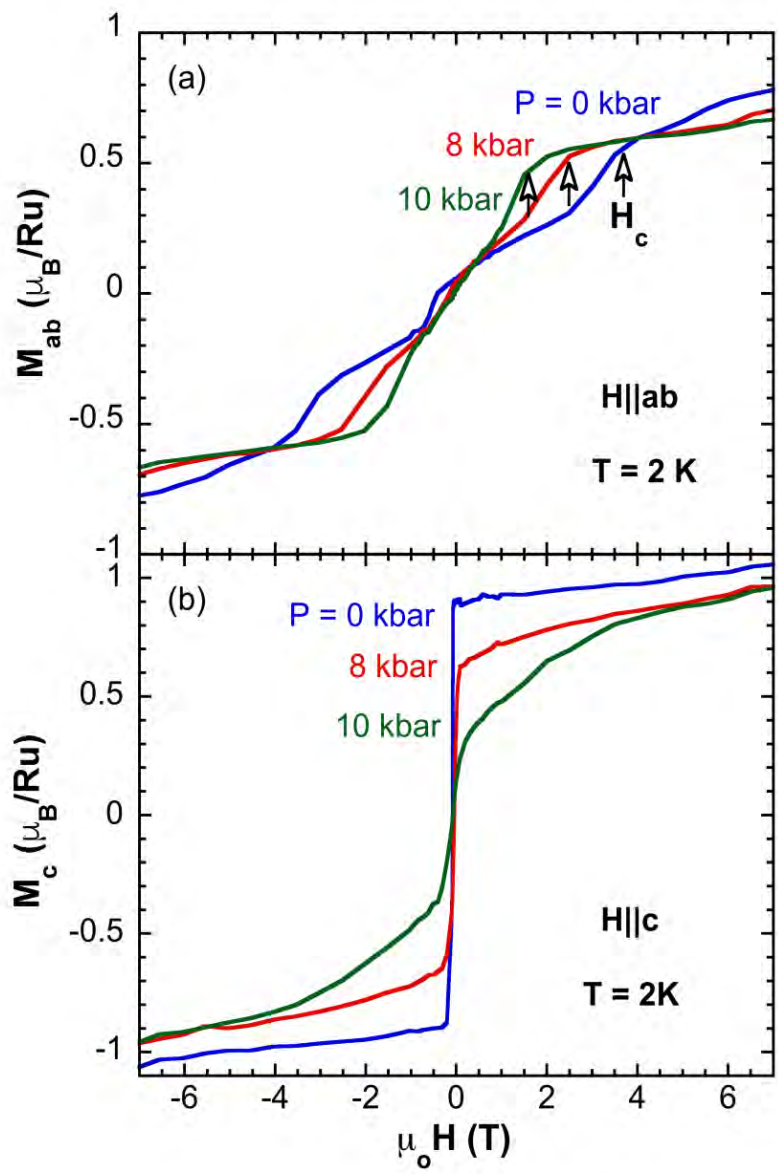

**Fig. 4**

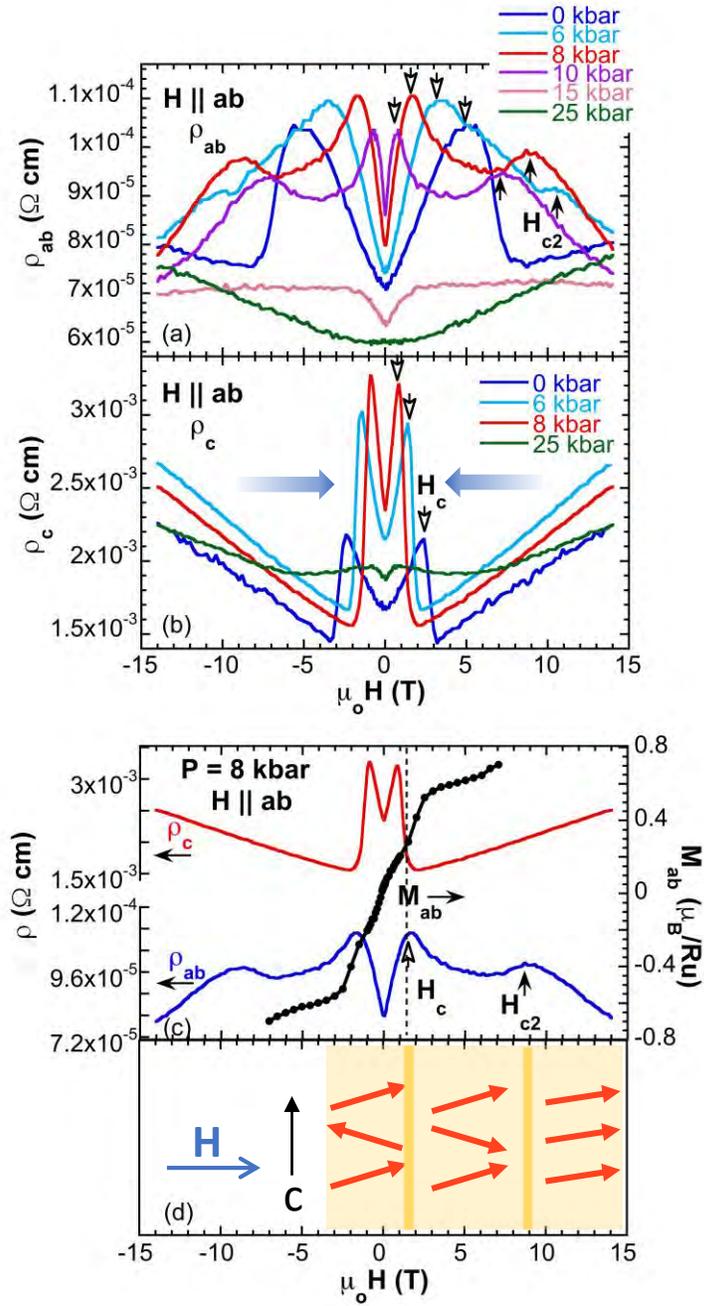

**Fig. 5**

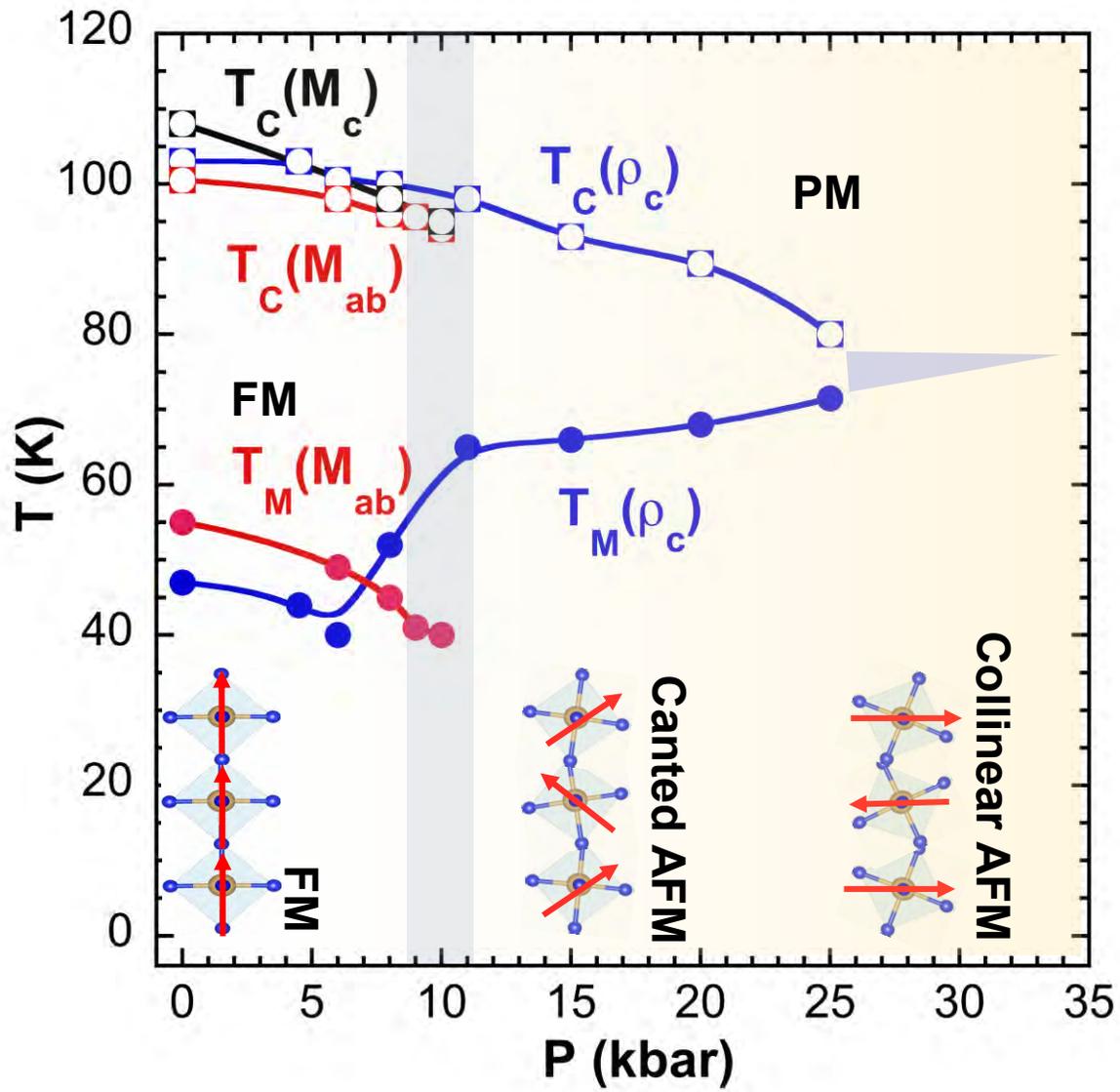

**Fig. 6**